**On-Chip Integrated Quantum-Dot Silicon-Nitride Microdisk Lasers**

*Weiqiang Xie, Thilo Stöferle, Gabriele Rainò, Tangi Aubert, Suzanne Bisschop, Yunpeng Zhu, Rainer F. Mahrt, Pieter Geiregat, Edouard Brainis, Zeger Hens, and Dries Van Thourhout\**

W. Xie, Y. Zhu, Prof. D. Van Thourhout
Photonics Research Group and Center for Nano- and Biophotonics (NB-Photonics), Ghent University, Technologiepark-Zwijnaarde 15 iGent, 9052 Ghent, Belgium
E-mail: Dries.VanThourhout@UGent.be

Dr. T. Stöferle, Dr. G. Rainò, Prof. R. F. Mahrt
IBM Research-Zurich, Säumerstrasse 4, 8803 Rüschlikon, Switzerland

Dr. T. Aubert, S. Bisschop, Dr. P. Geiregat, Prof. E. Brainis, Prof. Z. Hens
Physics and Chemistry of Nanostructures, Ghent University, Krijgslaan 281-S3, 9000 Ghent, Belgium

Photonic integrated circuits (PICs) offer new ways to manipulate and precisely control photons at the micro- and nanoscale. Mature manufacturing processes from the nanoelectronics industry have enabled PICs to become a cost-effective solution for high-volume applications in information processing and sensing. These circuits also increasingly provide the basis of breakthrough experiments in fundamental optics research on cavity quantum electrodynamics and quantum optics. In this respect, the silicon nitride (SiN) integrated photonics platform stands out because of its seamless transparency from the conventional telecom windows in the near infrared to the blue side of the visible spectrum, and low loss SiN PICs addressing all regions of this broad wavelength range have been demonstrated.[1-7]

Practical applications such as on-chip spectroscopy or biosensing require SiN PICs to be extended with active components, the most important of these being compact and tunable lasers. These could be realized through the combination of SiN waveguides with a suitable optical gain material. Colloidal quantum dots (QDs) offer unique opportunities in this respect. A variety of types of QDs and their 2D counterparts, colloidal quantum wells, have been shown in the past few years to exhibit optical gain at wavelengths that can be readily adjusted from near infrared to visible wavelengths through size quantization and material choice,[8-16] which provides for a perfect match with broadband SiN photonics. Various laser devices exploiting these colloidal gain media have been demonstrated, either using accidental *ad hoc* resonators,[17,18] vertical cavities in which gain materials are embedded between reflectors,[12-14,19] or individual cavities where QDs are coated on the surface.[20-29] Despite the confirmation of QDs as an optical gain medium in microlasers, the lack of a suitable integration process and limited QD stability have inhibited the development of more complex, fully integrated waveguide-coupled laser sources that could propel PICs into radically new applications.

Fabricating on-chip QD/SiN lasers requires an effective methodology to handle solution-processable QDs within a top-down CMOS-like manufacturing process. Such a process should preserve the optical properties of the QDs and facilitate a strong spatial overlap between the QDs and the resonating optical modes without degrading the optical quality factor ($Q$) of the hybrid device. Moreover, the QD/SiN resonators need to be coupled with low loss on-chip waveguides to facilitate efficient extraction and further manipulation of the emitted light. In these respects, the use of SiN membranes with locally embedded QD films within a CMOS-like process flow is a potent approach that allows for the combination of active QD/SiN resonators and passive SiN circuits on the same platform.[30,31]

Here, we use this approach to demonstrate a first on-chip QD/SiN microdisk laser coupled to planar SiN waveguides. The microdisk consists of a SiN/QD/SiN sandwich that supports high $Q$ whispering gallery modes (WGMs) with a maximum of optical confinement in the central QD layer. Lasing is achieved for different disk diameters under picosecond optical pumping. The lasing action is characterized via measuring the photoluminescence (PL) intensity versus pump power, and features a non-linear increase above a very low threshold fluence of 27 μJ·cm$^{-2}$ for a 7 μm disk. In addition, pronounced spectral narrowing, reduced emission lifetime and enhanced temporal coherence are observed above threshold. These ultra-compact waveguide-coupled QD/SiN microdisk lasers showcase the key building block for realizing active PICs on the SiN platform.

**Figure 1**a depicts the proposed device whereby a QD/SiN microresonator is vertically coupled to a passive SiN access waveguide. Design parameters, such as the disk diameter and thickness, the waveguide width, and the coupling offset and gap, are carefully chosen to enable a low lasing threshold and efficient waveguide coupling (see Supplementary Material S1). Finite-difference time-domain (FDTD) simulations (see Experimental Section) indicate a 7 μm × 0.25 μm monolithic SiN disk supports WGMs with a simulated $Q$ factor of ≈$5 \times 10^4$ at 625 nm, the central emission wavelength of the QDs to be integrated. Importantly, the optical mode is over two times more intense in the central section of the disk as compared to its top or bottom. Since the effective index of the QD film and the deposited SiN are similar (see Supplementary Material S2), the same enhancement is expected for the optical confinement in the QD layer of a SiN/QD/SiN sandwich.

The entire fabrication of the device is done using CMOS-like processes (see Experimental Section). First, we define the passive SiN waveguide layer, which is subsequently planarized through silicon oxide deposition and polishing. Subsequently, the (100/55/100 nm) SiN/QD/SiN layer stack is deposited and the microdisks are etched (see Supplementary Material S1). The embedded QDs are 9 nm diameter *flash* CdSe/CdS QDs with a ≈30 nm wide PL centered near 625 nm.[32] Figure 1b shows optical and scanning electron microscope (SEM) images of a fabricated disk array and a close-up of a selected single disk. Observation of these images reveals the presence of a well-defined circular boundary and flat top of the disk as well as the controlled position of the bus waveguide buried below the disk.

**Figure 2**a represents PL spectra of 5, 7 and 10 μm diameter disks acquired from the cleaved facet of the bus waveguide together with the background emission taken from the top of an unpatterned SiN/QD/SiN region. Whereas the latter aspect is in agreements with the featureless emission characteristic of the QD band-edge recombination, the spectra coupled out through the bus waveguides clearly show the WGMs of the disks with negligible background emission. The center of their envelope at ≈639 nm is redshifted by 15 nm with respect to the background emission spectrum. This shift is most likely due to reabsorption of emitted light in the QD layer, which occurs more frequently at shorter wavelengths (see Supplementary Material S2).

The WGM spectra consist of a series of transverse-electric (TE) and transverse-magnetic (TM)-like modes. Both mode families can be distinguished by their $Q$ factors and free spectral range (FSR), with TE modes exhibiting the larger FSR and higher $Q$ factors according to FDTD simulations (see Supplementary Material S3). An increase in $Q$ factor for the TE modes typically occurs when disk size and wavelengths increase, as can be seen in Figure 2b. The measured values remain well below the simulated ones, and we attribute this discrepancy to a combination of self-absorption (at shorter wavelengths) and scattering losses not accounted for in the simulations. TE mode scattering losses increase strongly with decreasing radius due to a stronger mode overlap with the disk edge. Assuming negligible self-absorption at 660 nm, we estimate at 630 nm the scattering loss decreases from 160 to 35 cm$^{-1}$ for a corresponding disk radius increasing from 5 to 15 μm diameter, as shown in Figure 2b. Performance of the gain medium must be sufficient to overcome such losses if the disk is to support lasing modes.

To assess quantitatively the optical gain characteristics of the CdSe/CdS QDs used here, we first determined the non-linear absorbance of a CdSe/CdS QD dispersion measured using transient absorption spectroscopy (see Supplementary Material S4). Following femtosecond pulsed photoexcitation at 520 nm, we observe optical gain in a 638-655 nm wavelength range at a pump pulse energy of 38.7 μJ·cm$^{-2}$. By further increasing the pump fluence, this initial gain turns into a broad gain band that extends from 600 to 674 nm with a material gain peaking at 930 cm$^{-1}$ at a wavelength of 623 nm. As population-inversion lasts for more than 80 ps and the concomitant transient absorbance decays at a rate of 4 ns$^{-1}$ (see Supplementary Material S4), we attribute optical gain in these QDs to stimulated emission from biexciton states, where interfacial alloying may somewhat suppress Auger recombination of the biexcitons.[18,33,34]

Next we determined the modal gain of waveguides defined in a SiN/QD/SiN stack identical to that used for realizing the microdisks. **Figure 3**a shows the optical emission of a waveguide defined by etching a SiN/QD/SiN stack similar to that used in the resonators and having dimensions of 600 μm in length and 5 μm in width. Upon pumping the waveguide using 400 nm femtosecond pulsed laser light focused by a cylindrical lens to a rectangular stripe, a broadband spontaneous emission spectrum is recorded from the cleaved edge of the waveguide. Increasing pump power corresponds to an observed marked narrowing and amplification of the spontaneous emission (ASE). The superlinear intensity increase saturates at a pump power ≈1.8 times above threshold power $P_{ASE}$ and the corresponding ASE spectrum is centered at ≈627 nm with a full-width at half-maximum (FWHM) of ≈8 nm. Consequently, we conclude that our process flow for the SiN/QD/SiN stacks preserves the gain characteristics of the CdSe/CdS QDs. In order to quantify this gain, we analyzed the emission from several SiN/QD/SiN waveguides with dissimilar lengths while maintaining a constant pump power density. Observation of Figure 3b reveals a sharp, superlinear rise in emission intensity for waveguides longer than ≈200 μm that saturates for lengths exceeding ≈400 μm. As seen in the microscope images, this ASE signal is perfectly guided by the waveguide without observable scattering, and thus corroborates the low intrinsic losses of the QD/SiN waveguides. We estimate the net modal gain $g$ by fitting the distance-dependent intensity to $I = A_0(e^{gl} - 1)/g$[35] within the region of exponential intensity increase. Averaged over several sets of nominally identical waveguides, net modal gains of around 70 cm$^{-1}$ and 100-120 cm$^{-1}$ are obtained for pump powers of about 1.4×$P_{ASE}$ and 2×$P_{ASE}$, respectively. The resulting numbers are in line with expectations. Considering a modal confinement of 23% and QD volume filling factor of ≈53%, these correspond to a material gain of up to 980 cm$^{-1}$, which is comparable to the material gain as deduced from the transient absorption spectra (see Supplementary Material S4). More importantly, lasing in QD/SiN microdisks should be feasible as the modal gain exceeds the aforementioned cavity losses.

To analyze the occurrence of lasing in the QD/SiN microdisks, we optically pump the disks with a pulsed laser at a wavelength of 400 nm (see Experimental Section). The pump light is coupled into a multi-mode fiber and then focused to a ≈12 μm diameter spot on the chip with a nearly flat-top intensity profile and 10 picoseconds pulse duration. The emission spectra of a 7 μm diameter disk under different excitation conditions are represented in **Figure 4**a as an example result (the laser characterization of larger disks can be found in Supplementary Material S5). Below the threshold ($0.89P_{th}$ – see below for a determination of $P_{th}$), the spectrum exhibits typical WGMs within the envelope of the broadband spontaneous emission of the colloidal QDs. By increasing the pump fluence above the threshold ($1.16P_{th}$), a sharp, 40-fold increase of the intensity for the WGM near 629 nm is observed. This is accompanied by line-narrowing from 0.58 nm to 0.14 nm FWHM, a distinct characteristic of the onset of lasing. Polarization analysis of the PL spectrum provides additional evidence that the lasing mode is a 1st order TE WGM (see Supplementary Material S6). By further increasing the pump intensity to $1.8P_{th}$, a second lasing mode appears at shorter wavelength (near 620 nm), consistent with the blue-shift of the gain spectrum with increasing of pump power (see Supplemental Material S4). The transition to lasing concurs with the emergence of scattered light of the WGMs in the PL image of the disk (see Insets of Figure 4a). In Figure 4b we plot the total output intensity versus the pump fluence, also known as light-in-light-out (L-L) curve. The laser threshold $P_{th}$ was determined to be $27 \pm 2$ μJ·cm$^{-2}$. Additionally, the log-scale L-L results for two lasing modes presented in the inset in Figure 4b are well-defined S-shaped curves, which can be accurately fitted by a static rate equation model (see Supplementary Material S7).

Stimulated emission is expected to shorten the luminescence decay time. We therefore performed time-resolved PL measurements, and Figure 4c shows the PL decay traces for different pump fluences. The extracted lifetimes dramatically decrease from a few nanoseconds below $P_{th}$ to tens of picoseconds above $P_{th}$, which is a value limited by the time resolution of the photon counter. Using a streak camera to precisely record the temporally and spectrally resolved dynamics, we find that the emission from the 7 μm disk lasts for about 7 ps and 11 ps (FWHM) for the short and long wavelength modes, respectively, see Figure 4d.

A key signature of lasing is high and extended coherence of the emitted photons. The degree of temporal coherence of the laser light is characterized by the first order correlation $g^{(1)}(\tau)$. By sending the light emitted from the waveguide end facets through a Michelson interferometer, we determine $g^{(1)}(\tau)$ from the interference fringe visibility $\gamma = |g^{(1)}(\tau)|$, where $\tau$ is now the time delay between two interferometer arms. We measure the interference pattern below and above threshold, and the results are reported in **Figure 5**. Below the lasing threshold, the envelope of the visibility can be appropriately fitted with a single exponential decay, resulting in a 1/e coherence time of $\tau_c$ = 0.33 ps. This very short coherence time equals the photon lifetime in the micro-resonator that can be calculated by $\tau_c = Q/\omega$, using the measured $Q$ of ≈1000 at 630 nm in 7 μm diameter disks. Above the lasing threshold, the coherence extends almost one order of magnitude to $\tau_c$ = 2.5 ps. This nearly equals the measured duration of the emitted laser pulse, but is slightly reduced due to a small temporal emission wavelength chirp from transients in the charge carrier density caused by the pulsed excitation scheme. The fringe pattern extends over the whole waveguide facet (see inset Figure 5b), as the single-transversal-mode design leads to perfect spatial coherence. A common feature both below and above threshold is that the multi-longitudinal-mode emission gives rise to a beat note where the peculiar ultrafast THz oscillation frequency corresponds to the frequency difference between the cavity modes (see inset Figure 5b). The almost Fourier-limited coherence, also in the multi-mode regime without noticeable mode competition, highlights the quality of the integrated QDs as excellent gain material.

The QD/SiN disk lasers preserve their properties over many weeks of measurements without significant degradation, wavelength drift or stability issues. This long-term stability is largely attributed to the efficient encapsulation of the colloidal QDs by the SiN matrix. Furthermore, the fabrication process is highly reproducible, allowing for a high device yield (>90%) while the operating wavelength has a variability of less than one nanometer for nominally identical devices.

We have created a versatile technology that enables hybrid integration of a whole class of solution-processable QDs with the SiN photonics platform. The quantitative analysis of the cavity and gain material allows for precise modeling and forecast of the actual device performance thanks to the stable fabrication process. Our device is the first waveguide-coupled colloidal QD laser and operates with an extremely low optical pump threshold of $P_{th}$= 27 µJ·cm$^{-2}$ in only 7 µm diameter disk at room temperature. We show a comprehensive characterization covering spectroscopic, temporal and coherence properties of these ultra-compact lasers. These results constitute a clear demonstration that wavelength-tunable, colloidal QDs can pave the way for versatile, active PICs for lab-on-a-chip, optofluidics, and sensing technologies. The achieved device stability together with excellent device-to-device and chip-to-chip reproducibility is critically important for high-volume fabrication and integration in practical applications. In future devices we expect that the threshold could be lowered even more by a further optimization of the core/shell QDs, and by switching from a top-pump to a waveguide-coupled pump, which would allow for extremely efficient, fully-integrated excitation schemes.[25] Furthermore, the insensitivity of colloidal QD gain to temperature[11] could be exploited to operate active PICs even in harsh environments.

**Experimental Section**

*Simulation and fabrication*: For simulation of the disk WGM, we used a freely available FDTD software package in cylindrical coordinates.[36] For device fabrication, all SiN-layers were deposited using a standard plasma-enhanced chemical vapor deposition (PECVD) system operating at a temperature of 270 °C. The PECVD-system is equipped with two RF frequency sources operating at respectively high-frequency (13.56 MHz) and low-frequency (100 – 460 kHz). All SiN layers were prepared using low-frequency, except for the top SiN layer of the SiN/QD/SiN sandwich, which was deposited using a mixed frequency mode to reduce the stress in the SiN-layer. Contact optical lithography was employed to define the SiN waveguide and disk patterns with photoresist as mask. The SiN and SiN/QD/SiN layers were etched with an anisotropic reactive ion etching process using a $CF_4/H_2$ gas mixture.

*Laser characterization*: The excitation light at a wavelength of $\lambda = 400$ nm is provided by a frequency-doubled regenerative amplifier seeded with a mode-locked Ti:Sapphire laser, resulting in an initial pulse duration of 100 – 200 fs with a repetition rate of 1 kHz. The light is then coupled into a multi-mode optical fiber with 25 μm core diameter and 100 cm length, which stretches the pulse duration to ≈10 ps and leads to homogenization of the beam profile towards a flat top. The output facet of the pump fiber is imaged demagnified at normal incidence onto the sample by microscope objectives, resulting in an approximately disk-shaped pump spot of 12 μm diameter, which intensity is controlled by a movable gradient filter. For the spectroscopic detection with a high-resolution spectrograph and the time-resolved measurements with a time-correlated single-photon counting system, the emitted light is collected with a multi-mode fiber (200 μm core, numerical aperture NA=0.22) directly from the cleaved waveguide end facet. For the interferometer, streak camera and polarization-dependent measurements, the light is collected with a microscope objective (NA=0.3) from the waveguide facet. In the Michelson interferometer, the emitted light is split with a non-polarizing beam-splitter cube, controllably delayed in one arm by a hollow retroreflector, which is mounted on a motorized linear stage, then recombined and focused on a cooled CCD.

**Supporting Information**

Supporting Information is available online from the Wiley Online Library or from the author.

**Acknowledgements**

The authors acknowledge the ERC-ULPPIC, H2020-MSCA phonsi and the IAP Photonics@be projects for financial support. The authors also thank Steven Verstuyft for help with optical index measurement.

Received: Sep 2016

[1] E. S. Hosseini, S. Yegnanarayanan, A. H. Atabaki, M. Soltani, A. Adibi, *Opt. Express* **2009**, 17, 14543.

[2] M. Ghulinyan, R. Guider, G. Pucker, L. Pavesi, *IEEE Photonics Technol. Lett.* **2011**, 23, 1166.

[3] J. F. Bauters, M. J. R. Heck, D. John, D. X. Dai, M. C. Tien, J. S. Barton, A. Leinse, R. G. Heideman, D. J. Blumenthal, J. E. Bowers, *Opt. Express* **2011**, 19, 3163.

[4] M. C. Tien, J. F. Bauters, M. J. R. Heck, D. T. Spencer, D. J. Blumenthal, J. E. Bowers, *Opt. Express* **2011**, 19, 13551.

[5] S. Romero-Garcia, F. Merget, F. Zhong, H. Finkelstein, J. Witzens, *Opt. Express* **2013**, 21, 14036.

[6] Q. Li, A. A. Eftekhar, M. Sodagar, Z. X. Xia, A. H. Atabaki, A. Adibi, *Opt. Express* **2013**, 21, 18236.

[7] A. Z. Subramanian, P. Neutens, A. Dhakal, R. Jansen, T. Claes, X. Rottenberg, F. Peyskens, S. Selvaraja, P. Helin, B. Du Bois, K. Leyssens, S. Severi, P. Deshpande, R. Baets, P. Van Dorpe, *IEEE Photonics J.* **2013**, 5.

[8] V. I. Klimov, A. A. Mikhailovsky, S. Xu, A. Malko, J. A. Hollingsworth, C. A. Leatherdale, H. J. Eisler, M. G. Bawendi, *Science* **2000**, 290, 314.

[9] R. D. Schaller, M. A. Petruska, V. I. Klimov, *J. Phys. Chem. B* **2003**, 107, 13765.

[10] V. I. Klimov, S. A. Ivanov, J. Nanda, M. Achermann, I. Bezel, J. A. McGuire, A. Piryatinski, *Nature* **2007**, 447, 441.

[11] I. Moreels, G. Raino, R. Gomes, Z. Hens, T. Stoferle, R. F. Mahrt, *Adv. Mater.* **2012**, 24, Op231.

[12] C. Dang, J. Lee, C. Breen, J. S. Steckel, S. Coe-Sullivan, A. Nurmikko, *Nat. Nanotechnol.* **2012**, 7, 335.

[13] B. Guzelturk, Y. Kelestemur, M. Olutas, S. Delikanli, H. V. Demir, *Acs Nano* **2014**, 8, 6599.

[14] J. Q. Grim, S. Christodoulou, F. Di Stasio, R. Krahne, R. Cingolani, L. Manna, I. Moreels, *Nat. Nanotechnol.* **2014**, 9, 891.

[15] C. X. She, I. Fedin, D. S. Dolzhnikov, A. Demortiere, R. D. Schaller, M. Pelton, D. V. Talapin, *Nano Lett.* **2014**, 14, 2772.

[16] C. X. She, I. Fedin, D. S. Dolzhnikov, P. D. Dahlberg, G. S. Engel, R. D. Schaller, D. V. Talapin, *Acs Nano* **2015**, 9, 9475.

[17] Y. Wang, V. D. Ta, Y. Gao, T. C. He, R. Chen, E. Mutlugun, H. V. Demir, H. D. Sun, *Adv. Mater.* **2014**, 26, 2954.

[18] Y. S. Park, W. K. Bae, T. Baker, J. Lim, V. I. Klimov, *Nano Lett.* **2015**, 15, 7319.

[19] B. Guzelturk, Y. Kelestemur, K. Gungor, A. Yeltik, M. Z. Akgul, Y. Wang, R. Chen, C. Dang, H. D. Sun, H. V. Demir, *Adv. Mater.* **2015**, 27, 2741.

[20] H. J. Eisler, V. C. Sundar, M. G. Bawendi, M. Walsh, H. I. Smith, V. Klimov, *Appl. Phys. Lett.* **2002**, 80, 4614.

[21] A. V. Malko, A. A. Mikhailovsky, M. A. Petruska, J. A. Hollingsworth, H. Htoon, M. G. Bawendi, V. I. Klimov, *Appl. Phys. Lett.* **2002**, 81, 1303.

[22] M. Kazes, D. Y. Lewis, Y. Ebenstein, T. Mokari, U. Banin, *Adv. Mater.* **2002**, 14, 317.

[23] S. I. Shopova, G. Farca, A. T. Rosenberger, W. M. S. Wickramanayake, N. A. Kotov, *Appl. Phys. Lett.* **2004**, 85, 6101.

[24] P. T. Snee, Y. H. Chan, D. G. Nocera, M. G. Bawendi, *Adv. Mater.* **2005**, 17, 1131.

[25] B. Min, S. Kim, K. Okamoto, L. Yang, A. Scherer, H. Atwater, K. Vahala, *Appl. Phys. Lett.* **2006**, 89, 191124.

[26] J. Schafer, J. P. Mondia, R. Sharma, Z. H. Lu, A. S. Susha, A. L. Rogach, L. J. Wang, *Nano Lett.* **2008**, 8, 1709.

[27] C. Grivas, C. Y. Li, P. Andreakou, P. F. Wang, M. Ding, G. Brambilla, L. Manna, P. Lagoudakis, *Nat. Commun.* **2013**, 4, 2376.

[28] M. M. Adachi, F. J. Fan, D. P. Sellan, S. Hoogland, O. Voznyy, A. J. Houtepen, K. D. Parrish, P. Kanjanaboos, J. A. Malen, E. H. Sargent, *Nat. Commun.* **2015**, 6, 8694.

[29] Y. Wang, K. E. Fong, S. C. Yang, V. D. Ta, Y. Gao, Z. Wang, V. Nalla, H. V. Demir, H. D. Sun, *Laser Photonics Rev.* **2015**, 9, 507.

[30] W. Q. Xie, Y. P. Zhu, T. Aubert, S. Verstuyft, Z. Hens, D. Van Thourhout, *Opt. Express* **2015**, 23, 12152.

[31] W. Q. Xie, Y. P. Zhu, T. Aubert, Z. Hens, E. Brainis, D. Van Thourhout, *Opt. Express* **2016**, 24, A114.

[32] M. Cirillo, T. Aubert, R. Gomes, R. Van Deun, P. Emplit, A. Biermann, H. Lange, C. Thomsen, E. Brainis, Z. Hens, *Chem. Mater.* **2014**, 26, 1154.

[33] G. E. Cragg, A. L. Efros, *Nano Lett.* **2010**, 10, 313.

[34] Y. S. Park, W. K. Bae, L. A. Padilha, J. M. Pietryga, V. I. Klimov, *Nano Lett.* **2014**, 14, 396.

[35] K. Shaklee, R. Nahory, R. Leheny, *J. Lumin.* **1973**, 7, 284.

[36] A. F. Oskooi, D. Roundy, M. Ibanescu, P. Bermel, J. D. Joannopoulos, S. G. Johnson, *Comput. Phys. Commun.* **2010**, 181, 687.

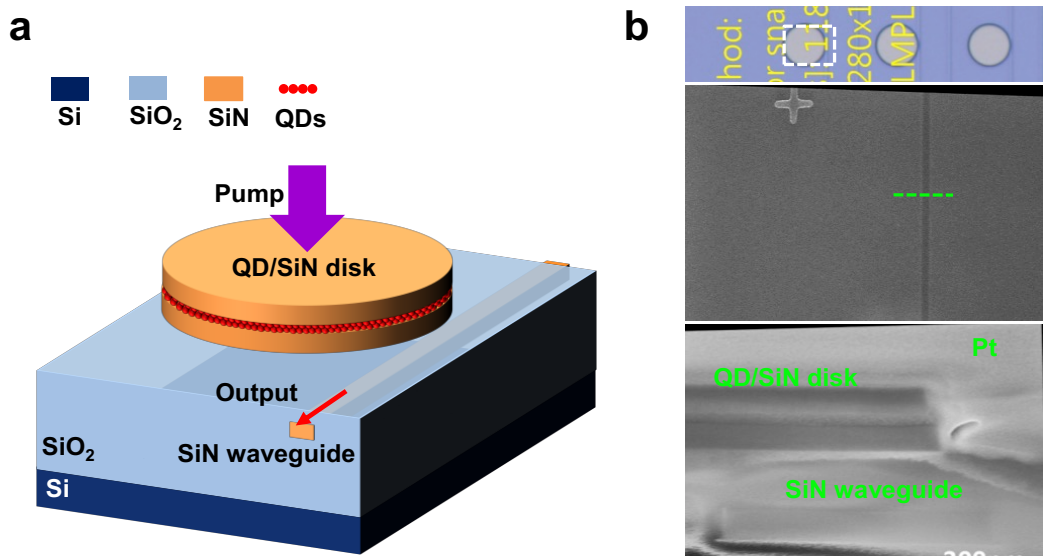

**Figure 1.** Device design and fabrication. a) Vertical coupling configuration of a QD/SiN disk and an access waveguide. A layer of QDs is embedded in the SiN disk. b) Optical microscope image and SEM image of a fabricated device. Top panel: optical photography of an array of devices on the chip. Middle: top view SEM image of a selected device. Bottom: Focused ion beam cross-sectional image of the waveguide-disk coupling region.

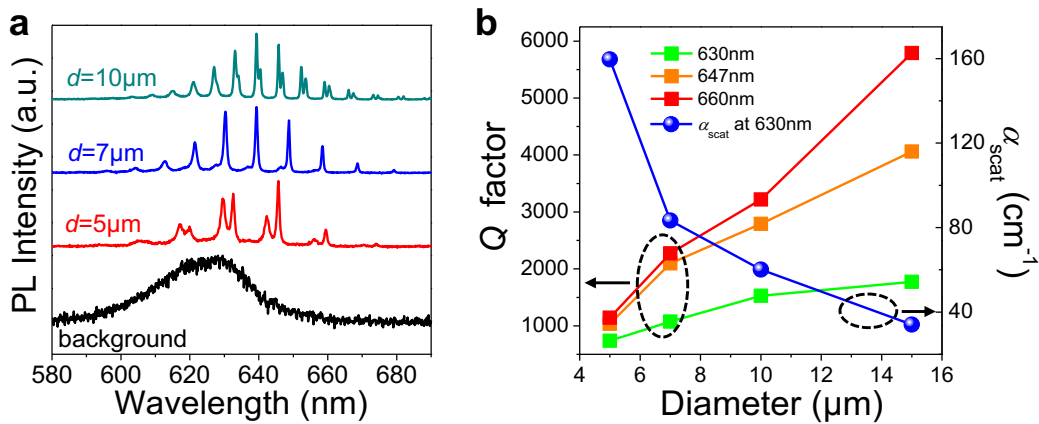

**Figure 2.** Characterization of disk WGMs at CW pump. a) Measured PL spectra of disks with different diameters together with background emission, excited by a CW 400 nm laser. b) Measured $Q$ factors for the 1$^{st}$ order TE WGMs near three selected wavelengths for disks of different diameters together with extracted scattering loss coefficients. Details of the analysis can be found in Supplementary Material S3.

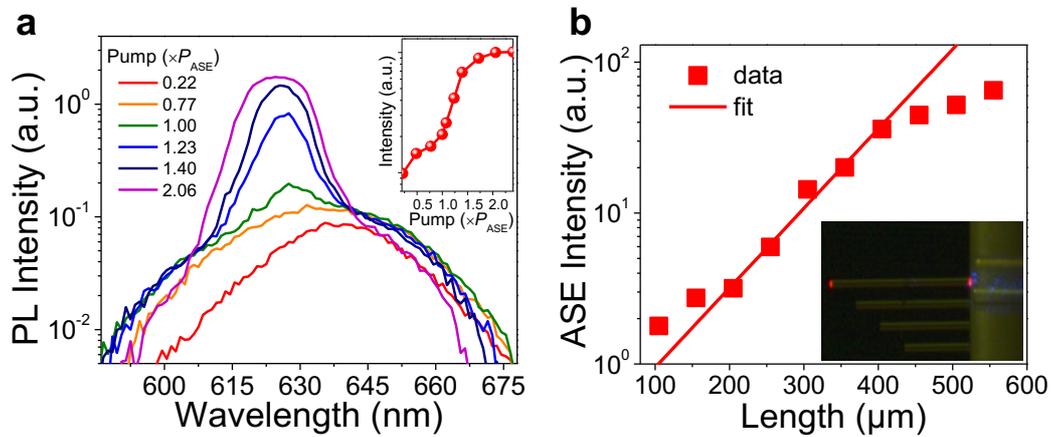

**Figure 3.** Optical gain in QD/SiN waveguide. a) The evolution of ASE spectra for ≈600 μm long SiN/QD/SiN waveguide when increasing the pump power. The latter is normalized to the pump power for which ASE starts dominating SE, $P_{ASE}$ (≈20 μJ·cm$^{-2}$). The integrated ASE intensity as a function of pump power is shown in the inset. b) ASE intensity versus the length of the waveguide at an excitation level of ≈2$P_{ASE}$. The inset shows the ASE signal in the waveguide is well-guided and appears as red emission from the waveguide's left and right end facets.

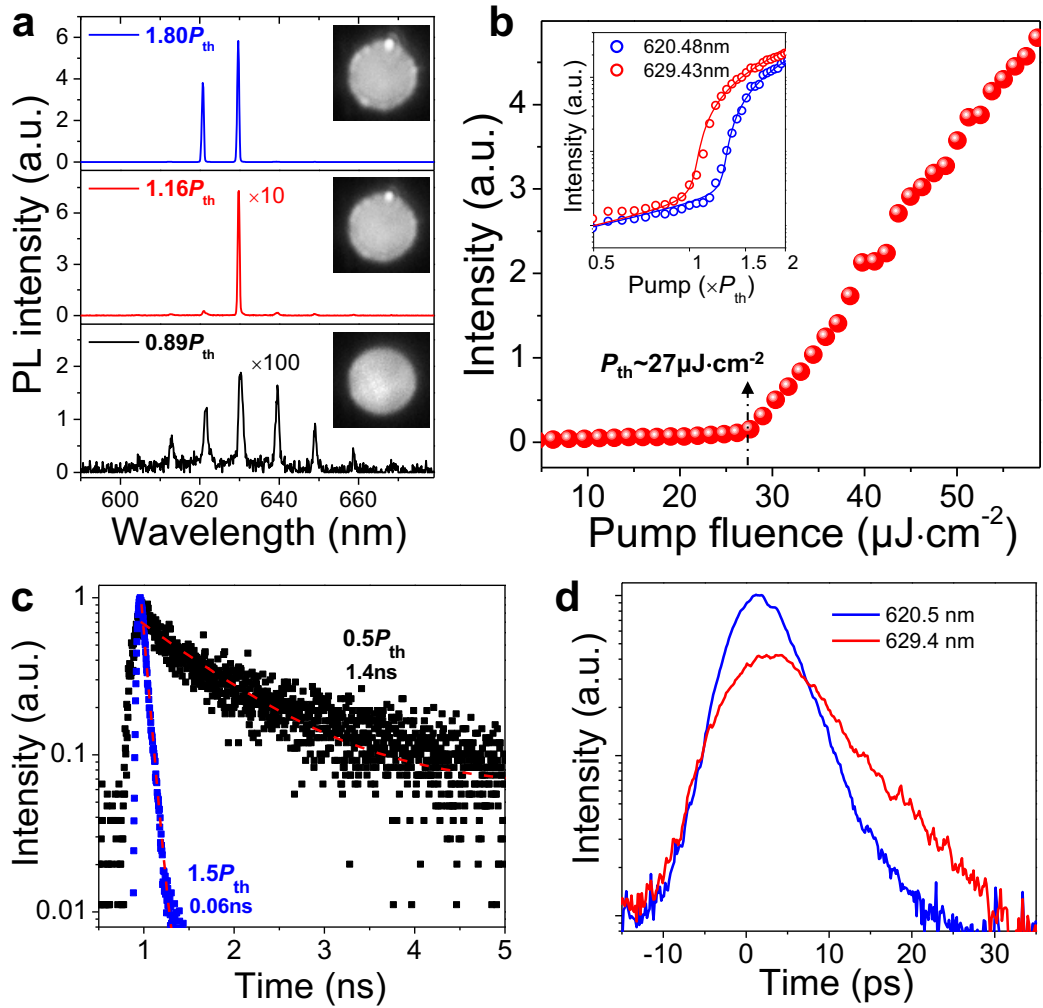

**Figure 4.** Laser threshold and lifetime characteristics. a) PL spectra of a 7 μm diameter disk under different pump fluences below and above the threshold $P_{th}$. Insets: Corresponding camera-recorded PL images of the disk (intensity normalized), showing the emergence of scattering from the WGMs above the threshold. b) Measured total PL intensity as a function of pump fluence, showing a clear threshold of 27 μJ·cm$^{-2}$. Inset: Log-scale Light(in)-Light(out) curves for two lasing modes. Symbols are measured data and solid lines are S-shaped curves obtained by a rate equation fit. c) Spectrally integrated decay traces at different pump fluences, together with extracted lifetimes from fitting a single-exponential-decay function as indicated by the red-dashed line. d) Temporal behavior of the two lasing WGMs at pump fluence of ≈3$P_{th}$.

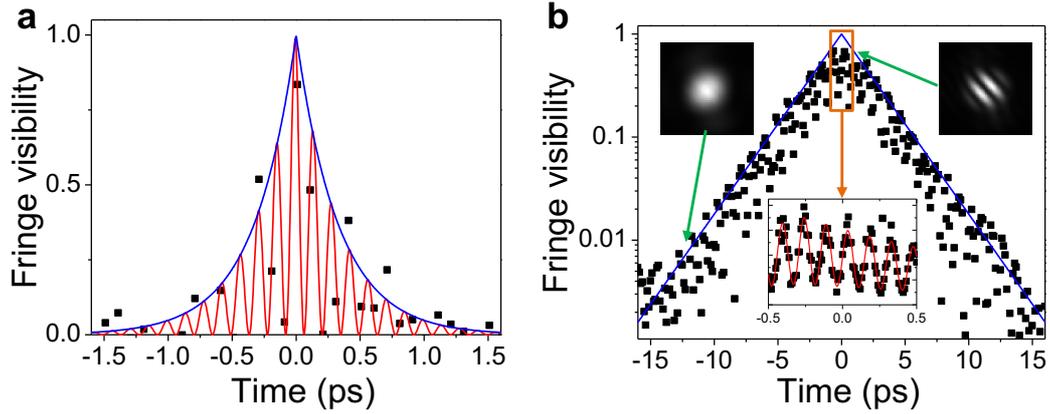

**Figure 5.** Temporal coherence of laser beam. a) Visibility of the interference fringes in a Michelson interferometer obtained below the lasing threshold (0.25$P_{th}$) as a function of delay time $\tau$ between the interferometer arms. The black symbols represent the measured data, the blue line shows a fitted envelope $\exp(-|\tau|/\tau_c)$ with $\tau_c$ = 0.33 ps and the red curve shows a fit taking into account the beating pattern of the two emitting cavity modes (see bottom inset of b). b) Above threshold (2.75$P_{th}$), the first order coherence lasts almost an order of magnitude longer, and a fit to the envelope yields $\tau_c$ = 2.5 ps (blue line). The top insets show exemplary interferograms of the emission from the waveguide end facet. The spread of the measured black data points is not noise but a consequence of the beating effect of multiple lasing modes, which is resolved when measuring with very high time resolution (bottom inset). The sine fit (red line) of the time-resolved beating pattern finds a period of 0.147 ps, corresponding to the inverse frequency mode spacing of the two lasing modes.